\documentclass[article]{aa}
\usepackage{graphicx}

\begin{document}
\title{The very local Hubble flow
\thanks{Based on observations made with the NASA/ESA Hubble Space
Telescope.  The Space Telescope Science Institute is operated by the
Association of Universities for Research in Astronomy, Inc. under NASA
contract NAS 5--26555.}}
\titlerunning{The very local Hubble flow }
\author{I. D. Karachentsev \inst{1}
\and M. E. Sharina \inst{1,10}
\and D. I. Makarov \inst{1,10}
\and A. E. Dolphin \inst{2}
\and E. K. Grebel \inst{3}
\and D. Geisler \inst{4}
\and P. Guhathakurta \inst{5}
\and P. W. Hodge \inst{6}
\and V. E. Karachentseva \inst{7}
\and A. Sarajedini \inst{8}
\and P. Seitzer \inst{9}}
\institute{Special Astrophysical Observatory, Russian Academy
of Sciences, N. Arkhyz, KChR, 369167, Russia
\and Kitt Peak National Observatory, National Optical Astronomy
Observatories,
P.O. Box 26732, Tucson, AZ 85726, USA
\and Max-Planck-Institut f\"{u}r Astronomie, K\"{o}nigstuhl 17, D-69117
Heidelberg, Germany
\and Departamento de F\'{\i}sica, Grupo de Astronom\'{\i}a, Universidad de
Concepci\'on,
Casilla 160-C, Concepci\'on, Chile
\and UCO/Lick Observatory, University of California at Santa Cruz, Santa
Cruz,
CA 95064, USA
\and Department of Astronomy, University of Washington, Box 351580,
Seattle,
WA 98195, USA
\and Astronomical Observatory of Kiev University, 04053, Observatorna 3,
Kiev,
Ukraine
\and Department of Astronomy, University of Florida, Gainesville, FL
32611,
USA
\and Department of Astronomy, University of Michigan, 830 Dennison
Building,
Ann Arbor, MI 48109, USA
\and Isaac Newton Institute, Chile, SAO Branch}
\date{Received:  December 2001}
%\maketitle
\abstract{
We present Hubble Space Telescope/WFPC2 images of eighteen galaxies
situated in the vicinity of the Local Group (LG) as part of an ongoing
snapshot survey of nearby galaxies. Their distances derived from the
magnitude of the tip of the red giant branch are 1.92$\pm$0.10 Mpc
(ESO 294-010), 3.06$\pm$0.37 (NGC~404), 3.15$\pm$0.32 (UGCA~105),
1.36$\pm$0.07
(Sex~B), 1.33$\pm$0.08 (NGC~3109), 2.64$\pm$0.21 (UGC~6817), 2.86$\pm$0.14
(KDG~90),
2.27$\pm$0.19 (IC~3104), 2.54$\pm$0.17 (UGC~7577), 2.56$\pm$0.15
(UGC~8508),
3.01$\pm$0.29 (UGC~8651), 2.61$\pm$0.16 (KKH~86), 2.79$\pm$0.26
(UGC~9240),
1.11$\pm$0.07 (SagDIG), 0.94$\pm$0.04 (DDO~210), 2.07$\pm$0.18 (IC~5152),
2.23$\pm$0.15
(UGCA~438), and 2.45$\pm$0.13 (KKH~98). Based on the velocity-distance
data
for 36 nearest galaxies around the LG, we find the radius of the
zero-velocity surface of the LG to be $R_0$ = (0.94$\pm$0.10) Mpc, which
yields
a total mass  $M_{LG} = (1.3\pm0.3)\cdot 10^{12} M_{\sun}$. The
galaxy distribution around
the LG reveals a Local Minivoid which does not contain any galaxy brighter
than
$M_V=-10$ mag  within a volume of $\sim100$ Mpc$^3$. The local Hubble flow
seems to be
very cold, having a one-dimensional mean random motion of $\sim30$ km s$^{-1}$.
The best-fit value of the local Hubble parameter is  73$\pm$15 km s$^{-1}$ 
Mpc$^{-1}$.
The luminosity function for the nearby field galaxies is far less steep
than one for members of the nearest groups.
%\end{abstract}
}
\maketitle

\section{Introduction}

Until quite recently the behaviour of the Hubble flow in the immediate
vicinity
of the Local Group (LG) remained practically unknown because of the
lack of reliable data on distances to many nearest galaxies. Lynden-Bell
(1981) and Sandage (1986) demonstrated that the velocity-distance relation
for galaxies around the LG should be non-linear due to gravitational
deceleration produced by the LG mass. Being determined from observational
data, the effect of deceleration permits calculation of the total mass of
the LG independently from mass estimates based on virial/orbital motions.
In the distance range of $(1 \div 3)$ Mpc, where the expected non-linearity
of the Hubble relation is significant, Sandage (1986) used
velocities and
distances of seven galaxies: Leo A (1.58 Mpc), IC 5152 (1.58), NGC 3109
(1.58), NGC 300 (1.66), Sex A (1.74), Sex B (1.74), and Pegasus (2.51 Mpc)
and estimated the LG mass to be $4\cdot10^{11} M_{\sun}$. As  became
clear later,
the distances of two of these seven galaxies (Leo A and Pegasus)
were
incorrect by more than a factor of 2. Based on largely the same
observational
data, Giraud (1986, 1990) obtained a three times higher mass for the LG.
Karachentsev \& Makarov (2001) used accurate distance measurements for
20 galaxies in the vicinity of the LG and determined the radius of the
zero-velocity sphere, $R_0$, which separates the LG from the total
cosmological expansion. From their data $R_0$ is of $0.96\pm0.05$ Mpc.
According
to Lynden-Bell (1981), this radius corresponds to the total LG mass,
$M_{LG} = (\pi^2/8G)\cdot H_0^2 \cdot R_0^3$, where $H_0$ is the local
Hubble parameter and  $G$  is the gravitational constant. For $H_0$ = 70
km s$^{-1}$ Mpc$^{-1}$
the radius of 0.96 Mpc yields a total mass of $1.2\cdot10^{12} M_{\sun}$.
Karachentsev
\& Makarov (2001) also noted that the dispersion of peculiar radial
velocities of galaxies around the LG is surprisingly small, only 
$\sim$25 km s$^{-1}$.

Over the last few years, rapid progress has been made in accurate
distance
measurements for nearby galaxies beyond the Local Group. At present, the
number of galaxies with known distances within $(1 \div 3)$ Mpc, is approaching
$\sim$40.
As a result of searching for new nearby objects on the POSS-II/ESO-SERC
plates (Karachentseva et al. 1998, 1999; Karachentsev et al. 2001a),
a number of low surface brightness dwarf galaxies have just recently been
found.
 The local galaxy complex candidates included in the program
of our snapshot survey with the Hubble Space Telescope (HST; programs GO-8192,
GO-8601, Seitzer et al. 1999).
We have reported the discovery of two very nearby dwarf galaxies,
KK 3 and KKR 25, situated within 2 Mpc (Karachentsev et al. 2000, 2001b).
Below we present new distance measurements for 18 more nearby galaxies
observed in the snapshot survey.

\section{ WFPC2 photometry and data reduction}

The galaxies were imaged with the Wide Field and Planetary Camera (WFPC2)
during 1999 October 15 to 2001 July 28 as  part of our HST snapshot survey
of nearby galaxy candidates (Seitzer et al. 1999). The
objects were usually centered on the WF3 chip, but for some bright
galaxies
the WFPC2 position was shifted towards the galaxy periphery to decrease
the
stellar crowding effect. 600-second exposures were taken in the F606W and
F814W filters for each object. Digital Sky Survey images (DSS-II, red) of
the
eighteen galaxies are shown in Figure 1 with the HST WFPC2 footprints
superimposed. The field size of the DSS images is 10$\arcmin \times
10\arcmin$.

 The photometric reduction was carried out using the HSTphot stellar
photometry package described by Dolphin (2000a). After removing
cosmic rays with the HSTphot \textit{cleansep} routine,
simultaneous photometry was performed on the F606W and F814W frames
using \textit{multiphot}, with aperture corrections for an aperture of
radius $0\farcs5$. Charge-transfer efficiency (CTE) corrections and
calibrations were then applied, which are based on the Dolphin (2000b)
formulae, producing $VI$ photometry for all stars detected in both images.
Because of the relatively small field of the Planetary Camera (PC) chip,
very few bright stars are available for the computation of the aperture
correction. Thus the PC photometry was omitted in further analysis.
Additionally, stars with the signal-to-noise ratio $S/N < 5 $,
$\mid \chi \mid\,\, >2.0$, or  $\mid$ sharpness $\mid\,\, >0.4$  in each
exposure were eliminated from the final photometry list.
We estimate the uncertainty of the photometric zeropoint to be within
$0\fm05$ (Dolphin 2000b).

  The WFPC2 images of the galaxies are presented in upper panels of
Figure 2, where both filters are combined. The compass in each field
indicates the North and East directions.

 \section{Color-Magnitude Diagrams and distances}

The middle panels of
Figure 2 show $I$ versus $(V-I)$ color-magnitude diagrams (CMDs) for the
eighteen observed galaxies. As demonstrated by Lee et al. (1993), the tip
of the red giant branch (TRGB) is a reliable distance indicator that is
relatively independent of age and metallicity. In the I band the TRGB for
low-mass stars is found to be stable within $ \sim 0.1$ mag (Salaris \& Cassisi,
1996; Udalski et al. 2001) for metallicities, [Fe/H], encompassing the
entire range from -2.1 up to -0.7 represented by Galactic globular clusters.
According to Da Costa \& Armandroff (1990), for metal-poor systems the TRGB
is located at $M_I = -4.05$ mag. Also, Lee et al. (1993) derived the TRGB to
be in the range of $-4.0 \pm0.1$ mag. Ferrarese et al. (2000) calibrated a
zero point of the TRGB from galaxies with Cepheid distances, and yielded
$M_I = -4.06 \pm0.07(random)\pm0.13(systematic)$ mag. A new TRGB calibration
$M_I = -4.04 \pm0.12$ mag was determined by Bellazzini et al.(2001) based on
photometry and a distance estimate from a detached eclipsing binary in the
Galactic globular cluster $ \omega$ Centauri. For this paper we use the
calibration of $M_I = -4.05$ mag.

   To determine the TRGB location, we obtained
a Gaussian-smoothed $I$-band luminosity
function for red stars with colors $V-I$ within $\pm0\fm5$ of
the mean $<V-I>$ for expected red giant branch (RGB) stars. Following
Sakai et al. (1996), we used a Sobel edge-detection filter.
The position of the TRGB
was identified with the peak in the filter response function. The
resulting
luminosity functions and the Sobel-filtered luminosity functions are
shown in the bottom panels of Figure 2.
The results are summarized in Table 1. The data listed in its columns
are as follow: (1) galaxy name; (2) equatorial coordinates corresponding
to the galaxy center; (3,4) apparent integrated magnitude and angular
dimension
from the NASA Extragalactic Database (NED); (5) heliocentric radial
velocity;
(6) morphological type in de Vaucouleurs notation; (7) position of the
TRGB
and its uncertainty derived with the Sobel filter; (8) Galactic extinction
in the  $I$- band (Schlegel et al. 1998); (9) true distance modulus
with its uncertainty, which takes into account the uncertainty in the TRGB
detection, as well as uncertainties of the HST photometry zero point
($\sim0\fm05$), the aperture corrections ($\sim0\fm05$), and crowding effects
($\sim0\fm06$) added in quadrature; the uncertainties in extinction and
reddening are taken to be $10\%$ of their values from Schlegel et al.(1998);
(more details on the total budget of internal and external systematic errors
for the TRGB method see in Mendez et al.(2002)); and
(10) linear distance in Mpc and its uncertainty. Some individual properties
of the galaxies are discussed below.

  {\em ESO 294-010 = PGC 01641.} The galaxy has a smooth regular shape and
the integrated color $(B-R)_T = 1.17$ (Jerjen et al. 1998), typical of
dwarf spheroidal (dSph) galaxies. These authors also measured the distance
to the galaxy, 1.71$\pm$0.07 Mpc, via fluctuations of surface brightness
and
found weak emission in [O\,{\sc iii}] and H$\alpha$ with a radial
velocity
of +117$\pm$5 km s$^{-1}$. Our CMD (Fig. 2) shows the presence
of a small number of faint blue stars situated in the central and southern
parts of the galaxy, which confirms the classification of the galaxy as a
dwarf system of mixed type, like Phoenix, LGS 3, and KK 3. Judging by its
velocity and distance, ESO 294-010 is a probable companion of the spiral
galaxy NGC 55.

  {\em NGC 404.} This compact lenticular galaxy with dusty patches in its
central part is partially affected by the halo of a bright star. Tonry et al. (2001)
determined its distance as 3.26$\pm$0.15 Mpc based on fluctuations of
its
surface brightness. NGC 404 is the nearest representative of the rare
class
of isolated S0 galaxies. There are no other known galaxies within 1.1 Mpc
around it (Karachentsev \& Makarov 1999). The origin of such an object is
unclear. This galaxy may represent the final stage of consecutive merging
of members of a former group of galaxies.

 {\em UGCA 105.} This irregular galaxy probably belongs to the
Maffei/IC342
group situated in a zone of strong Galactic extinction. Tikhonov et al.
(1992) and Karachentsev et al. (1997) resolved it into stars and
determined
its distance from the luminosity of the brightest stars to be 3.3
Mpc and 3.2 Mpc,
respectively. The distance, 3.15$\pm$0.32 Mpc, that we have derived
agrees well with these estimates.

  {\em Sextans B.} According to Sakai et al. (1997) the distance to this
dIrr galaxy found from the luminosity of its RGB stars
is 1.29$\pm$0.07 Mpc. Our estimate of its
distance, 1.36$\pm$0.07 Mpc, agrees well with the previous one.

  {\em NGC 3109.} Capaccioli et al. (1992) studied Cepheids in this galaxy
and derived from them a distance of 1.26$\pm$0.10 Mpc. Our distance
estimate
from the TRGB, 1.33$\pm$0.08 Mpc, agrees within the uncertainties with the
Cepheid distance.
According to Tully et al. (2002), NGC 3109 together with
three other dwarf galaxies, Sex~A, Sex~B, and Antlia, form a loose group
of
``squelched'' galaxies in a common dark halo.

  {\em UGC 6817 = DDO 99.} This dIrr galaxy has been resolved into stars
for
the first time by Georgiev et al. (1997), who determined its distance as
3.9$\pm$0.8 Mpc via the brightest stars. Our distance estimate from the TRGB
yields a lower distance, 2.64$\pm$0.21 Mpc, which corresponds to the
galaxy
location on the front side of the Canes Venatici cloud.

   {\em K 90 = DDO 113 = UGCA 276.} This dwarf galaxy of low surface
brightness
 and of regular smooth shape was resolved into stars by Makarova et al.
(1997).
 Jerjen et al. (2002) determined its integrated color $(B-R)$ = 1.07 mag and
the
central surface brightness  $\Sigma_B (0) = 24.92\pm0.08$ mag arcsec$^{-2}$. These
authors estimated the galaxy distance as 3.1$\pm$0.3 Mpc from surface brightness
fluctuations, which agrees well with our estimate, 2.86$\pm$0.14
Mpc,
from the TRGB. As one can see from its CMD (Fig. 2), K~90 does not contain a
significant number of blue stars, which favours this dwarf's
classification as a dSph
system. However, the galaxy shows strong HI emission, which makes us
regard it as a mixed dSph/dIrr system. Being situated 10$\arcmin$ away
from
the bright spiral galaxy NGC~4214, K~90 is a probable companion of the
galaxy because their velocities differ from each other by only 9
km s$^{-1}$.
Like UGC~6817, K~90 is situated on the front side of the CVn cloud.

  {\em IC 3104 = ESO 20-004 = PGC 039573.} As far as we know this
irregular
galaxy with a radial velocity of only $V_{LG}$ = 171 km s$^{-1}$ has now
been resolved into
stars for the first time. The distance, 2.27$\pm$0.19 Mpc, derived by us
is consistent with its low radial velocity. IC~3104 is a clear example of how
poorly
the immediate vicinity of the LG has so far been studied.

  {\em UGC 7577 = DDO 125.} This dIrr galaxy was resolved into stars by
Tikhonov \& Karachentsev (1998), who estimated its distance to be 4.8$\pm$1.0
Mpc
via the brightest stars. The TRGB position (Fig. 2) yields a much lower
distance, 2.54$\pm$0.17 Mpc, and allows us to consider UGC~7577 to be
situated
on the front side of the CVn cloud.

  {\em UGC 8508 = I Zw 60.} Karachentsev et al. (1994) resolved this dIrr
galaxy into stars and determined a distance of 3.7$\pm$0.8 Mpc from the
brightest stars. The TRGB position for UGC~8508 corresponds to a distance
of
2.56$\pm$0.15 Mpc, placing the galaxy in front of the M 101 group.

  {\em UGC 8651 = DDO 181.} The galaxy has a curved bow-like shape similar
to DDO 165. Its distance, 3.4$\pm$0.7 Mpc, was estimated by Makarova et
al.
(1998) via the brightest stars. The distance from its TRGB position is
3.01$\pm$0.29 Mpc. Like some above mentioned galaxies, UGC 8651
apparently belongs to the near side of the CVn cloud.

  {\em KKH 86.}  This dwarf irregular galaxy of low surface brightness
was found by Karachentsev et al. (2001a). Surface photometry of KKH~86 has
been
carried out by Makarova et al. (2002). The low TRGB distance of the
galaxy,
2.61$\pm$0.16 Mpc, agrees with its low velocity, $V_{LG}$ = 205 km s$^{-1}$.

  {\em UGC 9240 = DDO 190.} The galaxy was studied in detail by Aparicio
\& Tikhonov (2000), who derived a TRGB distance of 2.9$\pm$0.2 Mpc.
Our
WFPC2 photometry yields almost the same distance, 2.79$\pm$0.26 Mpc. The
galaxy is a probable member of the CVn cloud.

  {\em SagDIG.} This dIrr galaxy is situated at a low Galactic latitude.
Stellar photometry of SagDIG was made by Karachentsev et al. (1999) and
Lee \& Kim (2000), who found its TRGB distance to be 1.06$\pm$0.10 Mpc and
1.18$\pm$0.10 Mpc, respectively. Our determination of the distance,
1.04$\pm$0.05 Mpc, agrees well with both these estimates.

 {\em DDO 210 = Aquarius.} Detailed stellar photometry of the galaxy was
carried out by Lee et al. (1999), who derived the TRGB distance to be
0.95$\pm$0.05 Mpc. The WFPC2 photometry yields for DDO~210 the same
distance,
0.94$\pm$0.04 Mpc.

  {\em IC 5152.} This bright irregular galaxy contains a lot of blue stars
and dusty patches. Zijlstra \& Minniti (1998) detected its TRGB and
determined
the distance as 1.70$\pm$0.17 Mpc. Our distance estimate is 2.07$\pm$0.18
Mpc.
The lower value from Zijlstra \& Minniti  may be caused by crowding
effects,
higher for a ground-based telescope, which tends to brighten the
apparent
TRGB position (e.g., Stephens et al. 2001).

  {\em UGCA 438 = ESO 407 -018.}  According to Lee \& Byun (1999) the TRGB
distance of this irregular galaxy is 2.08$\pm$0.12 Mpc. Our WFPC2
photometry
gives a slightly higher value, 2.23$\pm$0.15 Mpc. The galaxy may be
a member
of the known loose group in Sculptor.

  {\em KKH 98.} This galaxy was found by Karachentsev et al. (2001a) as a
bluish
low surface brightness object partially resolved into stars. Its distance
derived from the TRGB is 2.45$\pm$0.13 Mpc. KKH~98 is a well-isolated galaxy.
There are no other known objects within 1 Mpc around it.

\section{The very local velocity field}

In addition to the 18 above-mentioned galaxies, there are 20 more
known
 galaxies which
are situated (or may be situated) within $D \sim 3$ Mpc around the LG
centroid.
A summary of these objects is given in Table 2.  Two galaxies from this
list, KK~3, Cetus have no measured velocities yet. Another two
southern galaxies, ESO~383-087 and IC~4662, have no distance estimates,
but their radial velocities with respect to the LG centroid are less than
150 km s$^{-1}$. For six more galaxies: SDIG, NGC~55, NGC~247, NGC~1569,
UGCA~92
and UGC~8638 their distances have been determined with low accuracy.
The data listed in the table columns are as follows: (1) galaxy name 
and cross-identification (lower line); (2)
equatorial coordinates (B1950.0); (3) integrated apparent magnitude (upper
line) and Galactic extinction from Schlegel et al. (1998) (lower line);
(4) angular dimension in arcmin (upper line) and morphological type in
de Vaucouleurs notation (lower line); (5) heliocentric radial
velocity and its standard
error in km s$^{-1}$; (6) distance to the galaxy and its standard error in Mpc;
(7) distance from the LG centroid, calculated as
    $$ D_{LG}^2 = D^2 + \Delta^2 - 2\cdot D\cdot\Delta\cdot\cos\theta,$$
where $\Delta$ = 0.44 Mpc is the distance of the Milky Way from the LG
centroid,
and $\theta$ is the angle between the direction to the galaxy and that to
M31;
(8) radial velocity of the galaxy with respect to the LG centroid with the
apex parameters  $V_\mathrm{a} =$316 km/s, $l_\mathrm{a}=$93$\degr$, $b_\mathrm{a}=$-4$\degr$
derived by Karachentsev \& Makarov (1996); (9) the above-
mentioned angle $\theta$; and (10) integrated absolute magnitude of the
galaxy.
The data of columns (2) -- (5) are taken from NED, and the sources of data
on
distances are given in footnotes to the table.

  The all-sky distribution of the 38 galaxies listed in Table 2
is presented in equatorial coordinates in Figure 3. The distribution
looks rather inhomogeneous, showing concentration of the objects
towards two opposite directions: the CVn cloud and the loose Sculptor
group,
which confirms the location of the Local Group in a filament, extending
from
Canes Venatici to Sculptor (Tully, 1988). Another feature of the
distribution
is complete absence of galaxies in a wide area around the direction
towards
Orion ($\alpha$ =$05^h40^m$, $\delta$ = $-3\degr$). This empty region, occupying
about a quarter
of the entire sky, is indicated by the right-hand solid curve in
Fig. 3. We nickname it as
a Local Minivoid (LMV). Unlike the Local Void (Tully, 1988), which has the
center position about $\alpha$ = $18^h38^m$, $\delta$ = $+18\degr$, an angular
diameter
$\sim60\degr$ and a depth of 1500 km s$^{-1}$ (Karachentseva et al. 1999), the
LMV
extends in depth to  $D \sim 5$ Mpc. In the
about 100 Mpc$^3$, there is not a single known galaxy brighter than
 $M_\mathrm{V} = -11^m$. An absence of galaxies in the direction of Orion may be partially
due to extinction by the Milky Way. However, the presence of such a
``perfect'' empty volume can not be explained by Galactic extinction
alone,  because some more distant dwarf galaxies ("Orion", A0554+07, UGC 3303,
kk49) with radial velocities of $V_\mathrm{LG} \simeq 400$ km/s are seen behind
the LMV (Karachentsev \& Musella, 1996).
  Velocities and distances of 34 galaxies from Table 2 with respect to the
LG centroid are shown in Figure 4 with indication of their observational
uncertainties.
As was demonstrated by Sandage (1986, 1987), the deceleration of
the cosmic
expansion in the vicinity of the LG leads to a curvature of the Hubble
relation,
which intersects the line of $V_\mathrm{LG} = 0$ at a non-zero distance $R_0$
from the LG
centroid. The value of $R_0$ characterizes the radius of the zero-velocity
surface that separates the LG from the total cosmic expansion.
In the case of spherical symmetry with the cosmological parameter $\Lambda
= 0$
the LG mass depends on $R_0$ as $M_\mathrm{LG} = (\pi^2/8G)\cdot H_0^2 \cdot
R_0^3$,
which yields for $H_0$ = 70 km s$^{-1}$ Mpc$^{-1}$ a mass  $M_\mathrm{LG}/M_{\sun} =
1.39\cdot10^{12}\cdot(R_0/{\rm Mpc})^3$. In Fig. 4 the thick solid line
shows
the velocity-distance regression with two parameters, $H_0$ = 73 km s$^{-1}$ 
Mpc$^{-1}$
and  $M_\mathrm{LG} = 1.26\cdot10^{12} M_{\sun}$, derived as the least-square
solution. The two thin solid lines correspond to the 95\% confidence
interval
of the regression.

   As one can see from these data, the majority of nearby galaxies
follow the Hubble relation very well. However, a group of 7 galaxies with
$D_\mathrm{LG} \sim2.8$ Mpc and $V_\mathrm{LG} \sim$270 km s$^{-1}$ departs appreciably from
the relation as a whole. All these objects are situated on the front side
of the Canes Venatici cloud and, apparently, move from us toward the cloud
center at an additional velocity of about 85 km s$^{-1}$. Another galaxy,
NGC~6789,
has the radial velocity much lower than the expected velocity. This object
is situated in front of the Local Void at a high Supergalactic latitude
($+41.6\degr$). The low velocity of NGC~6789 may be caused by the
gravitational action of the Local Void, which imparts a peculiar
velocity of $\sim$80 km s$^{-1}$ towards us on NGC~6789. Fitting the Hubble relation with its
two dependent parameters, $H_0$ and $M_{LG}$, we omitted the seven
probable
members of the CVn cloud (indicated as asterisks) and also two galaxies,
SDIG and NGC 247 (shown as triangles) due to their poorly determined 
distances.
For the remaining 27 galaxies the least-squares  method yields a
radius of
the zero-velocity surface of 0.94$\pm$0.10 Mpc. The mean-root-square
dispersion of radial velocities for field galaxies relative to the Hubble
regression with the indicated parameters is  $\sigma_v = 34$ km s$^{-1}$. A
significant
part of the scatter of galaxies in Fig. 4 is caused by the uncertainties
of the measured
distances, $\Delta D/D \sim$ 10\%, while measurement errors of the radial
velocities
are relatively small, $\sim$2\%. Taking the errors of the
observables into account, 
the mean-root-square random velocity of nearby galaxies
themselves is reduced to 
$\sigma_\mathrm{v}$ = 29 km s$^{-1}$.

  To get a complete picture of the local velocity field, we also show in
Fig. 4 the members of the Local Group, which are indicated by open
circles.
Observational data on their distances and velocities are listed in Table
3,
the columns of which contain: (1) galaxy name; (2) morphological type, (3)
measured
distance and its standard error; (4) heliocentric velocity and its error;
(5, 6) distance and radial velocity with respect to the LG centroid. The
last
two quantities are determined in the same manner as described above. The
basic data are taken from the book of van den Bergh (2000) and also from
Evans et al. (2000). Some of the other recent sources of distances are mentioned
in the table footnotes.

  Considering the structure and kinematics of two nearby groups around M81
(Karachentsev et al. 2002a) and Centaurus~A (Karachentsev et al. 2002b),
we noted a general similarity of these groups with the LG. The systems of
companions around M81 and Cen~A are characterized by a mean
projected
linear radius of $\sim$200 Kpc and  a radial velocity dispersion of
$\sim$80 km s$^{-1}$
each. For the companions of the Milky Way and M31, these quantities are
of the same order. As a robust estimate of mass we used the relation
$$M_\mathrm{orb} = (32/3\pi)\cdot G^{-1}\cdot (1- 2e^2/3)^{-1} <R_p\cdot \Delta
V^2_r
>,$$
which is valid for arbitrarily-oriented Keplerian orbits  with
eccentricity
$e$. Adopting $e$ = 0.7 as the average eccentricity, we derived the
following
mass estimates: $M$(M81) = $1.6\cdot10^{12} M_{\sun}$, and  $M$(CenA)
= $2.1\cdot10^{12} M_{\sun}$.
For the galaxies M31 and Milky Way the mass estimates from orbital motions
of their companions are $1.0\cdot10^{12} M_{\sun}$ (M31) and
$1.0\cdot10^{12}
M_{\sun}$ (Milky Way).
In the case of the Milky Way it was taken into account that we use the actual
spatial
distances to the companions, $D$, instead of projected distances, $R_p$.
It
should also be noted that a significant contribution to the Milky Way mass
estimate is made by only one remote companion, Phoenix, which has a
high peculiar velocity (Gallart et al. 2001). One can assume that the
orbit
of Phoenix does not fit the Keplerian motion because of its probable
location in the ``infall zone'' (see Fig. 4). Then, removing Phoenix
drops the Milky Way mass estimate to $0.5\cdot10^{12} M_{\sun}$,
which seems to us  more appropriate. The derived  orbital masses are quite
close to more sophisticated estimates of masses of the Milky Way and M31
inside $ \sim 50$ Kpc: $M_\mathrm{MW} = 0.4 \cdot10^{12} M_{\sun}$ (Fich \& Tremaine, 1991),
$M_\mathrm{MW} = (0.5\pm0.1)\cdot10^{12} M_{\sun}$ (Kochanek, 1996), $M_\mathrm{M31} =
0.8\cdot10^{12} M_{\sun}$ (Evans \& Wilkinson, 2000), $M_\mathrm{M31} =
(0.8\pm0.1)\cdot10^{12} M_{\sun}$ (C\^ot\'e et al. 2000), and $M_\mathrm{M31} =
(0.7 \div 1.0)\cdot10^{12} M_{\sun}$ (Evans et al. 2000). Therefore, the sum
of orbital (virial)
masses of both the LG subgroups, $(1.2 \div 1.5)\cdot10^{12} M_{\sun}$, agrees
within the errors with the total mass of the LG, $(1.3\pm0.3)\cdot10^{12}
M_{\sun}$,
derived from the radius of the zero-velocity surface. The agreement of
these mutually independent mass estimates  corresponding to essentially
different scale lengthes, $ \sim $ 50 kpc and $ \sim 1$ Mpc, tell us that the major part
of mass in the LG is strongly concentrated towards its two biggest members.

 \section {Concluding remarks}

  The total view of the Local Group and its surroundings is presented in
Figure 5 in the Cartesian Supergalactic coordinates. The upper panel shows
the Local Complex of galaxies seen in the Supergalactic plane. The nearby
galaxies are represented as ``illuminated balls'', whose size is proportional
to the galaxy luminosity. Spiral and irregular galaxies are indicated as
``dark balls'', and elliptical and spheroidal ones are shown as ``light
balls''.
The bottom panel presents the Local Complex from another point of view.
The Canes Venatici cloud is situated on the right side, and the Sculptor
group is located on the left side. The Local Minivoid occupies the bottom
left corner, and the Local Void of Tully extends above the complex plane
in the ``+Z''-direction. Only one galaxy, NGC~6789, is highly raised above
the rest of the objects. The morphological segregation is evident: the dEs
and
gas-deficient dSphs are tightly concentrated around the large spirals.

  Another manifestation of the morphological segregation can be seen from
the data of Figure 6, which presents the luminosity function (LF) of
nearby
galaxies. The upper histogram shows the absolute magnitude distribution
for 38 "field" galaxies from Table 2. The bottom one corresponds to
the combined LF of 96 galaxies in the LG, M81 group and CenA group. E and
dSph galaxies on the histograms are shaded. Because the effective depth
of both the samples is almost the same, we have the first opportunity here
to compare LFs of the field galaxies and the members of groups, avoiding
strong bias effects. Comparison of the two histograms confirms the well
known fact that in dense regions the LF has a steep shape
(Binggeli et al. 1985, Ferguson \& Sandage, 1990). However, this
difference
seems to be true for the integrated LF only, when all morphological
types of galaxies are combined. Thus, the median absolute
magnitude of dSph+dE galaxies is approximately the same ($\sim -10\fm7$)
for the field galaxies as well as for the members of the groups. The
median
absolute magnitudes of dIrr+S galaxies are also similar ($\sim-14\fm0$)
for
both the samples. Because the relative number of dSph + dE galaxies among
the field objects is only 16\%, and their fraction among the group members
is 53\%, the known effect of morphological segregation of galaxies
leads to the observed difference in their LFs. Therefore, considering or
ignoring gas-deficient dSphs may be the reason for a significant
difference
in LFs from one sample to another. For instance, the ``blind'' HIPASS survey
(Banks et al. 1999) is insensitive to dSph galaxies, which renders the
luminosity function of the HIPASS  sample less steep than for
optical surveys.

  One can assume that the LF of field galaxies is the ``initial'' LF, being
almost undistorted by interactions. In such a case the observed
difference in LF for members of groups with respect to the field LF may be
described as a result of two processes: a) formation of objects of high
luminosity (``cannibals'') due to merging of galaxies in groups, b) creation
of new "tidal" dwarfs, say, via fragmentation of tidal bridges and tails
(Duc et al. 2000).

  As it has been emphasized by Governato et al (1997) and Klypin et
al.(2002),
the dispersion of random motions of field galaxies and the centers of
groups,
$\sigma_v$, contains key information about the scenario of galaxy
formation
and the value of the average density of matter, $\Omega_m$. Sandage et al.
(1972)
noted that the velocity dispersion around the local Hubble flow is low
($\sigma_v \sim 70$ km s$^{-1}$), which corresponds to $\Omega_m \sim0.1$. New
observational
data, based on much more accurate distance measurements, yield an
essentially
lower value of $\sigma_\mathrm{v}$. Karachentsev et al. (2002a) showed that the
peculiar
velocity of the centroid of the M81/NGC~2403 group does not exceed 35
km s$^{-1}$.
According to Karachentsev et al. (2002a), peculiar radial velocities of the
centroids of the CenA and the M83 groups are $+18\pm24$ km s$^{-1}$ and
$-17\pm27$ km s$^{-1}$,
respectively. From analysis of the local velocity field Karachentsev \&
Makarov (2001) found the value and the direction of the solar apex with
respect
to the Local Volume galaxies with $D < 7$ Mpc:  $V_\mathrm{a} = (325\pm11)$ km s$^{-1}$,
$l_\mathrm{a} = (95\pm2)\degr, b_\mathrm{a} = (-4\pm1)\degr$,
where $l$ and $b$ are the galactic coordinates. Because the apex of
solar motion with respect to the LG centroid is $V_a =
316$ km s$^{-1}$, $l_a = 93\degr$, and $b_a = -4\degr$, the peculiar velocity of
the LG
centroid with respect to the whole Local Volume turns out to be only about
10 km s$^{-1}$.

  Therefore, we conclude that the centroids of all nearby well-studied
groups have surprisingly low dispersion of peculiar velocities, which is
$\sigma_v \sim (20 \div 30)$ km s$^{-1}$. The same value is also characteristic of the
field
galaxies in the vicinity of the LG. Baryshev et al. (2001) suggested that
the observed quiescence of the local Hubble flow is a signature of the
cosmological vacuum-dominated universe where the velocity
perturbations
are adiabatically decreasing.

\acknowledgements
 We thank the referee,John Huchra, for his useful comments.
Support for this work was provided by NASA through grant GO--08601.01--A
from
the Space Telescope Science Institute, which is operated by the
Association
of Universities for Research in Astronomy, Inc., under NASA contract
NAS5--26555.
This work was partially supported by
 RFBR grant 01--02--16001 and DFG-RFBR grant 01--02--04006.
 D.G. acknowledges financial support for
this project received from CONICYT through Fondecyt grant 8000002.

 The Digitized Sky Surveys were produced at the Space Telescope
Science Institute under U.S. Government grant NAG W--2166. The
images of these surveys are based on photographic data obtained
using the Oschin Schmidt Telescope on the Palomar Mountain and the UK
Schmidt Telescope. The plates were processed into the present
compressed digital form with permission of these institutions.

 This project made use of the NASA/IPAC Extragalactic Database (NED),
which is operated by the Jet Propulsion Laboratory, Caltech, under
contract
with the National Aeronautics and Space Administration.

{}

\newpage
\onecolumn
\begin{table}
\caption{New distances to very nearby galaxies around the Local Group}
\begin{tabular}{llrrrrrrrr}\\ \hline
 Name     & $\alpha$ (1950.0) $\delta$ &  $B_T$  &  $a \times b$  & $V_h \pm\sigma$
&$ T$ & $I(TRGB)$& $A_I$ &$(m-M)_0$ & $D$  \\
  & $^{hh mm ss}\;\;\;\;\;   \degr\degr \;\arcmin\arcmin \;\arcsec\arcsec$
&  mag  &   arcmin   & km s$^{-1}$   &   &  mag   & mag & mag   &Mpc \\ \hline
ESO294$-$010& 002406.2$-$420755 & 15.60 & 1.1$\times$0.7 & 117$\pm$5 &
$-$1:& 22.38 & 0.01& 26.42 &1.92\\
PGC 01641 &                 &       &         &        &   &   0.05 &
&   .10 & .10\\
NGC 404   & 010639.2 352705 & 11.21 & 3.5$\times$3.5 & $-$48$\pm$9 & $-$1&
23.49 & 0.11& 27.43 &3.06\\
  &                 &       &         &        &   &    .24 &     &   .25
& .37\\
UGCA 105  & 050935.6 623431 & 13.9  & 5.5$\times$3.5 & 111$\pm$5 & 10&
24.05 & 0.61& 27.49 &3.15\\
  &                 &       &         &        &   &    .20 &     &   .22
& .32\\
Sex B     & 095723.1 053421 & 11.85 & 5.1$\times$3.5 & 301$\pm$1 & 10&
21.72 & 0.06& 25.66 &1.36\\
DDO 070   &                 &       &         &        &   &    .09 &
&   .13 & .07\\
NGC 3109  & 100049.0$-$255504 & 10.39 &19.7$\times$3.7 & 403$\pm$1 & 10&
21.70 & 0.13& 25.62 &1.33\\
  &                 &       &         &        &   &    .10 &     &   .13
& .08\\
UGC 6817  & 114816.8 390930 & 13.4  & 4.1$\times$1.5 & 242$\pm$1 & 10&
23.11 & 0.05& 27.11 &2.64\\
DDO 099   &                 &       &         &        &   &    .15 &
&   .17 & .21\\
KDG 090   & 121227.1 362948 & 15.40 & 1.5$\times$1.3 & 284$\pm$6 & $-$1:&
23.27 & 0.04& 27.28 &2.86\\
DDO 113   &                 &       &         &        &   &    .06 &
&   .11 & .14\\

IC 3104   & 121545.0$-$792654 & 13.63 & 3.8$\times$1.8 & 430$\pm$5 &  9&
23.49 & 0.76& 26.78 &2.27\\
  &                 &       &         &        &   &    .16 &     &   .18
& .19\\
UGC 7577  & 122515.4 434613 & 12.84 & 4.3$\times$2.4 & 196$\pm$4 & 10&
23.01 & 0.04& 27.02 &2.54\\
DDO 125   &                 &       &         &        &   &    .10 &
&   .13 & .17\\
UGC 8508  & 132847.1 551002 & 14.40 & 1.7$\times$1.0 &  62$\pm$5 & 10&
23.02 & 0.03& 27.04 &2.56\\
I Zw060   &                 &       &         &        &   &    .09 &
&   .13 & .15\\
UGC 8651  & 133744.2 405931 & 14.7  & 2.3$\times$1.3 & 201$\pm$1 & 10&
23.35 & 0.01& 27.39 &3.01\\
DDO 181   &                 &       &         &        &   &    .19 &
&   .21 & .29\\
KKH 86    & 135202.2 042917 & 16.8  & 0.7$\times$0.5 & 283$\pm$3 & 10&
23.08 & 0.05& 27.08 &2.61\\
  &                 &       &         &        &   &    .10 &     &   .13
& .16\\
UGC 9240  & 142248.4 444504 & 13.25 & 1.8$\times$1.6 & 150$\pm$4 & 10&
23.20 & 0.02& 27.23 &2.79\\
DDO 190   &                 &       &         &        &   &    .18 &
&   .20 & .26\\
SagDIG    & 192705.4$-$174659 & 14.12 & 2.9$\times$2.1 & $-$77$\pm$4 & 10&
21.26 & 0.22& 25.09 &1.04\\
  &                 &       &         &        &   &    .03 &     &   .10
& .05\\
DDO 210   & 204406.5$-$130155 & 14.0  & 2.2$\times$1.1 &$-$137$\pm$5 & 10&
20.91 & 0.10& 24.86 &0.94\\
Aquarius  &                 &       &         &        &   &    .05 &
&   .10 & .04\\
IC 5152   & 215926.6$-$513214 & 11.06 & 5.2$\times$3.2 & 124$\pm$3 & 10&
22.58 & 0.05& 26.58 &2.07\\
  &                 &       &         &        &   &    .16 &     &   .18
& .18\\
ESO407$-$018& 232347.3$-$323950 & 13.86 & 1.5$\times$1.2 &  62$\pm$5 & 10&
22.72 & 0.03& 26.74 &2.23\\
UGCA 438  &                 &       &         &        &   &    .12 &
&   .15 & .15\\
KKH 98    & 234303.9 382624 & 16.7  & 1.1$\times$0.6 &$-$136$\pm$3 & 10&
23.14 & 0.24& 26.95 &2.45\\
  &                 &       &         &        &   &    .06 &     &   .11
& .13\\
\end{tabular}
\end{table}
\small
\begin{table}
\caption{Properties of very nearby galaxies at the Local Group boundaries and around the Local Group}
\begin{tabular}{lcrcccllcrrr}\\ \hline
 Name    & $\alpha$ (1950.0) $\delta$ &$B_T$ & $a \times b$ & $V_h$& $\sigma$&  $D $&
$\sigma$ &$D_{LG}$ &$V_{LG}$& $\theta $&$M_B$ \\
&                 &$A_B$ &  $T$    &      &    &      &     &         &
&          &       \\
\hline
SDIG   & 000541.0$-$345124 &15.48&  1.1$\times$0.9 & 207& 7& 1.6 & .8 &
1.56& 216 & 76.2 &$-$10.59\\
E349$-$031 &                 & 0.05&    10    &    &  &     &    &     &
&      &      \\
NGC 55   & 001238.0$-$392954 & 8.84& 32.4$\times$5.6 & 129& 3& 1.66&  .4&
1.65& 111 &  80.7& $-$17.32\\
&                 & 0.06&     9    &    &  &     &    &     &     &      &
\\
KK 3     & 001300.0$-$322736 &14.85&  1.3$\times$1.0 &   $-$&  & 1.92&
.19& 1.85&  $-$  & 73.7 &$-$11.63\\
E410$-$005 &                 & 0.06&    $-$3    &    &  &     &    &     &
&      &      \\
Cetus    & 002338.7$-$111916 &14.4 &  5.0$\times$4.3 &   $-$&  & 0.78&
.05& 0.62&  $-$  & 52.4 &$-$10.18\\
KKSG 01  &                 & 0.12&    $-$2    &    &  &     &    &     &
&      &      \\
E294$-$010 & 002406.2$-$420755 &15.60&  1.1$\times$0.7 & 117& 5& 1.92&
.10& 1.92&  81 & 83.2 &$-$10.84\\
PGC01641 &                 & 0.02&    $-$1:   &    &  &     &    &     &
&      &      \\
NGC 247  & 004439.6$-$210200 & 9.86& 21.4$\times$6.9 & 160& 2& 2.48& .6 &
2.31& 215 & 62.0 &$-$17.19\\
&                 & 0.08&     7    &    &  &     &    &     &     &      &
\\
NGC 300  & 005231.8$-$375715 & 8.95& 21.9$\times$15.5& 144& 1& 2.1 & .2 &
2.06& 114 & 79.0 &$-$17.72\\
&                 & 0.06&     7    &    &  &     &    &     &     &      &
\\
NGC 404  & 010639.2 352705 &11.21&  3.5$\times$3.5 & $-$48& 9& 3.06& .37&
2.62& 195 &  7.6 &$-$16.47\\
&                 & 0.25&    $-$1    &    &  &     &    &     &     &
&      \\
NGC 1569 & 042604.6 644423 &11.86&  3.6$\times$1.8 &$-$104& 4& 2.5 & .5 &
2.18&  88 & 39.5 &$-$18.15\\
&                 & 3.02&     9    &    &  &     &    &     &     &      &
\\
UGCA 92  & 042727.1 633025 &13.8:&  2.0$\times$1.0 & $-$99& 5& 1.8 & .5 &
1.49&  89 & 39.5 &$-$15.90\\
&                 & 3.42&    10    &    &  &     &    &     &     &      &
\\
UGCA 105 & 050935.6 623431 &13.9 &  5.5$\times$3.5 & 111& 5& 3.15& .32&
2.85& 279 & 44.3 &$-$14.94\\
&                 & 1.35&    10    &    &  &     &    &     &     &      &
\\
Leo A    & 095632.4 305910 &12.92&  5.1$\times$3.1 &  20& 4& 0.69& .06&
0.87& $-$44 & 98.7 &$-$11.36\\
DDO 069  &                 & 0.09&    10    &    &  &     &    &     &
&      &      \\
Sex B    & 095723.1 053421 &11.85&  5.1$\times$3.5 & 301& 1& 1.36&
.07& 1.63& 111 &120.4 &$-$13.96\\
DDO 070  &                 & 0.14&    10    &    &  &     &    &     &
&      &      \\
NGC 3109 & 100049.0$-$255504 &10.39& 19.7$\times$3.7 & 403& 1& 1.33& .08&
1.71& 110 &143.9 &$-$15.52\\
&                 & 0.29&    10    &    &  &     &    &     &     &      &
\\
Antlia   & 100147.0$-$270521 &16.19&  2.0$\times$1.5 & 361& 9& 1.32& .06&
1.70&  65 &144.7 & $-$9.75\\
PGC029194&                 & 0.34&    10    &    &  &     &    &     &
&      &      \\
Sex A    & 100829.5$-$042646 &11.86&  5.9$\times$4.9 & 324& 1&
1.42& .08& 1.74&  94 &130.1 &$-$14.09\\
DDO 075  &                 & 0.19&    10    &    &  &     &    &     &
&      &      \\
UGC 6817 & 114816.8 390930 &13.4 &  4.1$\times$1.5 & 242& 1& 2.64& .21&
2.74& 248 & 99.0 &$-$13.82\\
DDO 099  &                 & 0.11&    10    &    &  &     &    &     &
&      &      \\
KDG 090  & 121227.1 362948 &15.40&  1.5$\times$1.3 & 284& 6& 2.86& .14&
2.98& 286 &102.7 &$-$11.87\\
DDO 113  &                 & 0.09&    $-$1:   &    &  &     &    &     &
&      &      \\
NGC 4214 & 121308.2 363620 &10.24&  8.5$\times$6.6 & 291& 3& 2.94& .18&
3.07& 295 &102.2 &$-$17.19\\
&                 & 0.09&     9    &    &  &     &    &     &     &      &
\\
\end{tabular}
\end{table}
\begin{table}
\begin{tabular}{lcrcccllcrrr}\\ \hline
IC 3104  & 121545.0$-$792654 &13.63&  3.8$\times$1.8 & 430& 5& 2.27& .19&
2.63& 171 &141.4 &$-$14.85\\
&                 & 1.70&     9    &    &  &     &    &     &     &      &
\\
UGC 7577 & 122515.4 434613 &12.84&  4.3$\times$2.4 & 196& 4& 2.54& .17&
2.62& 241 & 95.2 &$-$14.27\\
DDO 125  &                 & 0.09&    10    &    &  &     &    &     &
&      &      \\
GR 8     & 125610.9 142914 &14.68&  1.1$\times$1.0 & 214& 3& 2.1 & .2 &
2.38& 136 &124.4 &$-$12.04\\
DDO 155  &                 & 0.11&    10    &    &  &     &    &     &
&      &      \\
UGC 8508 & 132847.1 551002 &14.40&  1.7$\times$1.0 &  62& 5& 2.56& .15&
2.55& 186 & 83.3 &$-$12.70\\
I Zw060  &                 & 0.06&    10    &    &  &     &    &     &
&      &      \\
UGC 8638 & 133658.5 250144 &14.47&  1.2$\times$0.8 & 274& 4& 2.3 & .5 &
2.50& 273 &112.7 &$-$12.40\\
&                 & 0.06&    10    &    &  &     &    &     &     &      &
\\
UGC 8651 & 133744.2 405931 &14.7 &  2.3$\times$1.3 & 201& 1& 3.01& .29&
3.10& 271 & 97.0 &$-$12.72\\
DDO 181  &                 & 0.03&    10    &    &  &     &    &     &
&      &      \\
E383$-$087 & 134623.0$-$354848 &11.03&  4.5$\times$3.6 & 326& 1&  $-$  &
&  $-$  & 108 &166.1 &      \\
PGC049050&                 & 0.30&     8    &    &  &     &    &     &
&      &      \\
KKH 86   & 135202.2 042917 &16.8 &  0.7$\times$0.5 & 283& 3& 2.61& .16&
2.92& 205 &131.6 &$-$10.40\\
&                 & 0.12&    10    &    &  &     &    &     &     &      &
\\
KK 230   & 140501.5 351809 &17.9 &  0.6$\times$0.5 &  61& 1& 1.9 & .2 &
2.03& 125 &101.3 & $-$8.55\\
&                 & 0.06&    10    &    &  &     &    &     &     &      &
\\
DDO 187  & 141338.6 231713 &14.38&  1.7$\times$1.3 & 154& 4& 2.5 & .2 &
2.70& 174 &112.2 &$-$12.71\\
UGC 9128 &                 & 0.10&    10    &    &  &     &    &     &
&      &      \\
UGC 9240 & 142248.4 444504 &13.25&  1.8$\times$1.6 & 150& 4& 2.79& .26&
2.84& 266 & 91.3 &$-$14.03\\
DDO 190  &                 & 0.05&    10    &    &  &     &    &     &
&      &      \\
KKR 25   & 161237.3 542946 &16.45&  1.1$\times$0.9 &$-$135& 2& 1.86& .18&
1.79&  72:& 74.3 & $-$9.94\\
	 &                 & 0.04&    $-$1:   &    &  &     &    &     &
&      &      \\
IC 4662  & 174212.0$-$643718 &11.74&  2.8$\times$1.6 & 308& 4&  $-$  &
&  $-$  & 145 &132.3 &      \\
&                 & 0.30&     9    &    &  &     &    &     &     &      &
\\
NGC 6789 & 191617.1 635250 &13.76&  1.3$\times$1.0 &$-$141& 9& 3.6 & .20&
3.33& 144 & 50.1 &$-$14.32\\
&                 & 0.30&    10    &    &  &     &    &     &     &      &
\\
SagDIG   & 192705.4$-$174659 &14.12&  2.9$\times$2.1 & $-$77& 4& 1.04&
.05& 1.15&  23 & 93.0 &$-$11.60\\
&                 & 0.49&    10    &    &  &     &    &     &     &      &
\\
DDO 210  & 204406.5$-$130155 &14.0 &  2.2$\times$1.1 &$-$137& 5& 0.94&
.04& 0.94&  13 & 76.7 &$-$11.09\\
Aquarius &                 & 0.22&    10    &    &  &     &    &     &
&      &      \\
\end{tabular}
\end{table}
\begin{table}
\begin{tabular}{lcrcccllcrrr}\\ \hline

IC 5152  & 215926.6$-$513214 &11.06&  5.2$\times$3.2 & 124& 3& 2.07& .18&
2.19&  75 & 98.9 &$-$15.63\\
&                 & 0.11&    10    &    &  &     &    &     &     &      &
\\
Tucana   & 223827.3$-$644054 &15.7 &  2.9$\times$1.2 & 130: & 2& 0.88& .04&
1.10&   9: &108.3 & $-$9.16\\
PGC069519&                 & 0.14&    $-$2    &    &  &     &    &     &
&      &      \\
E407$-$018 & 232347.3$-$323950 &13.86&  1.5$\times$1.2 &  62& 5& 2.23&
.15& 2.17&  99 & 75.7 &$-$12.94\\
UGCA 438 &                 & 0.06&    10    &    &  &     &    &     &
&      &      \\
KKH 98   & 234303.9 382624 &16.7 &  1.1$\times$0.6 &$-$136& 3& 2.45& .13&
2.02& 152 & 11.2 &$-$10.78\\
&                 & 0.53&    10    &    &  &     &    &     &     &      &
\\
WLM      & 235924.4$-$154422 &11.03& 11.5$\times$4.0 &$-$116& 2& 0.92&
.04& 0.78& $-$10 & 57.5 &$-$13.95\\
DDO 221  &                 & 0.16&     9    &    &  &     &    &     &
&      &      \\
\hline
\multicolumn{1}{l}{SDIG} &\multicolumn{11}{l}{Laustsen et al. (1977)}
\\
\multicolumn{1}{l}{NGC 55}&  \multicolumn{11}{l}{Puch\'e \& Carignan (1988)}
\\
\multicolumn{1}{l}{KK 3}&\multicolumn{11}{l}{Karachentsev et al. (2000)}\\
\multicolumn{1}{l}{Cetus}&\multicolumn{11}{l}{Sarajedini et al. (2001)}\\
\multicolumn{1}{l}{NGC 247}&\multicolumn{11}{l}{Puch\'e \& Carignan (1988),
Tammann (1987)}\\
\multicolumn{1}{l}{NGC 300}& \multicolumn{11}{l}{Freedman et al. (1992),
Soffner et al. (1996)}\\
\multicolumn{1}{l}{NGC 1569}& \multicolumn{11}{l}{Karachentsev et al.
(1994), Waller (1991)}\\
\multicolumn{1}{l}{UGCA 92}&  \multicolumn{11}{l}{Karachentsev et al.
(1994)}\\
\multicolumn{1}{l}{Leo A}&   \multicolumn{11}{l}{Tolstoy et al. (1998)}\\
\multicolumn{1}{l}{Antlia}&\multicolumn{11}{l}{Aparicio et al. (1997),
$D=1.57\pm0.07$ Mpc, Castellani et al.(2001)} \\
\multicolumn{1}{l}{Sex A}&  \multicolumn{11}{l}{Sakai et al. (1996)} \\
\multicolumn{1}{l}{NGC 4214}&  \multicolumn{11}{l}{Maiz$-$Apellaniz et
al.(2002), $D=2.67\pm.20$, Drozdovsky et al.(2001a)}\\
\multicolumn{1}{l}{GR 8}&  \multicolumn{11}{l}{Dohm$-$Palmer et al.
(1998)}\\
\multicolumn{1}{l}{UGC 8638}&   \multicolumn{11}{l}{Makarova et al.
(1998)}\\
\multicolumn{1}{l}{KK 230}&  \multicolumn{11}{l}{Grebel \& Guhathakurta
(2001)}\\
\multicolumn{1}{l}{DDO 187}& \multicolumn{11}{l}{Aparicio et al. (2000)}
\\
\multicolumn{1}{l}{KKR 25}&  \multicolumn{11}{l}{Karachentsev et al.
(2001b);the heliocentric velocity of $-$135 km s$^{-1}$}\\
&     \multicolumn{11}{l}{is probably caused by the Local HI}\\
\multicolumn{1}{l}{NGC 6789}&   \multicolumn{11}{l}{Drozdovsky et
al.(2001b)}\\
\multicolumn{1}{l}{Tucana}&      \multicolumn{11}{l}{Oosterloo et al.
(1996);
the heliocentric velocity of 130 km s$^{-1}$}\\
&     \multicolumn{11}{l}{is probably caused by the Local HI}\\
\multicolumn{1}{l}{WLM}& \multicolumn{11}{l}{Dolphin (2000c)} \\ \hline

\end{tabular}
\end{table}

\begin{table}
\caption{Distances and radial velocities of the LG members}
\begin{tabular}{lrcrrrcr} \\ \hline
 Name    & Type &     $ D$     &      $V_h\pm\sigma$   &  $D_{LG}$ &
$V_{LG}$ \\
&      &     Mpc    &     km s$^{-1}$  &  Mpc  &   km s$^{-1}$ \\  \hline
IC 10    &  10  &  0.83  .05 &   $-$344  5 &  0.44 &   $-$60  \\
NGC 147  &  $-$3  &  0.73  .05 &   $-$193  3 &  0.30 &    85  \\
And III  &  $-$3  &  0.76  .04 &   $-$355 10 &  0.32 &   $-$92  \\
NGC 185  &  $-$3  &  0.62  .03 &   $-$202  7 &  0.19 &    73  \\
NGC 205  &  $-$5  &  0.82  .04 &   $-$244  3 &  0.38 &    24  \\
M 32     &  $-$5  &  0.80  .04 &   $-$205  3 &  0.36 &    61  \\
M 31     &   3  &  0.78  .04 &   $-$301  1 &  0.34 &   $-$35  \\
And I    &  $-$3  &  0.81  .04 &   $-$380  2 &  0.37 &  $-$120  \\
SMC      &  10  &  0.06  .01 &    158  4 &  0.47 &   $-$22  \\
Sculptor &  $-$3  &  0.08  .01 &    110  1 &  0.43 &    96  \\
LGS 3    &  $-$1: &  0.62  .02 &   $-$286  4 &  0.25 &   $-$74  \\
IC 1613  &  10  &  0.73  .02 &   $-$232  5 &  0.48 &   $-$89  \\
And V    &  $-$3  &  0.81  .05 &   $-$403  4 &  0.38 &  $-$143  \\
And II   &  $-$3  &  0.53  .11 &   $-$188  3 &  0.12 &    46  \\
M 33     &   5  &  0.81  .05 &   $-$180  1 &  0.40 &    36 \\
Phoenix  &  $-$1: &  0.44  .03 &    $-$52  6 &  0.61 &  $-$145 \\
Fornax   &  $-$3  &  0.14  .01 &     53  2 &  0.44 &   $-$33 \\
LMC      &   9  &  0.05  .005&    278  2 &  0.47 &    28 \\
Carina   &  $-$3  &  0.10  .005&    223  3 &  0.50 &   $-$53 \\
Leo I    &  $-$3  &  0.25  .03 &    285  2 &  0.59 &   128  \\
Sex dSph &  $-$3  &  0.086 .004&    226  1 &  0.50 &     8   \\
Leo II   &  $-$3  &  0.21  .01 &     76  1 &  0.56 &   $-$18    \\
Ursa Min &  $-$3  &  0.066 .003&   $-$247  1 &  0.42 &   $-$44   \\
Draco    &  $-$3  &  0.086 .009&   $-$293  1 &  0.41 &   $-$48   \\
Sgr dSph &  $-$3  &  0.024 .002&    142  1 &  0.45 &   164   \\
NGC 6822 &  10  &  0.49  .04 &    $-$57  2 &  0.65 &    63   \\
Cas dSph &  $-$3  &  0.71  .04 &   $-$307  2 &  0.31 &    $-$5   \\
Pegasus  &  10  &  0.76  .10 &   $-$182  2 &  0.45 &    61  \\
Peg dSph &  $-$3  &  0.78  .04 &   $-$354  3 &  0.40 &   $-$93 \\
Milky Way&   4  &  0.01  .002&      0 10 &  0.44 &   $-$16 \\
\hline
\multicolumn{1}{l}{LGS 3:}& \multicolumn{5}{l}{Miller et al. (2002)} \\
\multicolumn{1}{l}{Phoenix:}& \multicolumn{5}{l}{Gallart C., et al.
(2001)}\\
\multicolumn{1}{l}{Cas dSph:}& \multicolumn{5}{l}{mistake in l,b$-$
coordinates}\\
& \multicolumn{5}{l}{(Evans et al. 2000) is corrected}\\
\multicolumn{1}{l}{IC1613:}& \multicolumn{5}{l}{Dolphin et al. (2001)}\\
\hline
\end{tabular}
\end{table}
\clearpage
\begin{figure}
%\centering
%\includegraphics[width=12cm]{Fig1.ps}
%\vspace{5mm}

\caption{ Digital Sky Survey images of 18 galaxies in the vicinity of
the Local Group. The field size is 10$\arcmin$, North is up and East is
left. The HST WFPC2 footprints are superimposed.}
%\end{figure*}
%\clearpage

%\begin{figure*}
%\centering
%\vspace{-10mm}

%\includegraphics[width=16.5cm]{Fig2_3_4+1.ps}
%\vspace{-18mm}

\caption{{\bf Top}: WFPC2 images of 18 galaxies: ESO 294-010, NGC~404, UGCA~105,
 Sex~B,
NGC~3109, UGC~6817, KDG~90, IC~3104, UGC~7577, UGC~8508, UGC~8651, KKH~86,
UGC~9240, SagDIG, DDO~210, IC~5152, UGCA~438, and KKH~98 produced by
combining the two 600s exposures obtained through the F606W and F814W
filters. The arrows point to the North and the East.
{\bf Middle}: The color-magnitude diagrams from the WFPC2 data for
the 18 galaxies around the Local Group.
{\bf Bottom}: the Gaussian-smoothed $I$-band luminosity function restricted
to red stars (top), and the
output of an edge-detection filter applied to the luminosity function
for the 18 very nearby galaxies studied here.}

\caption{ All-sky distribution of the 38 galaxies around the LG in
equatorial coordinates. The Local Void (LV) of Tully (1988) and the
Local Minivoid (LMV) are outlined with solid lines.}
%\end{figure*}

%\begin{figure*}
%\includegraphics[width=12.0cm]{Fig6.ps}
%\vspace{5mm}
\caption{ Velocity-distance relation from the data in Table 2 (filled circles).
Velocities and distances have been corrected to the centroid of the LG.
The members of the LG from Table 3 are shown as open circles. Observational
1-$\sigma$ errors are indicated with bars. The thick solid line shows
the velocity-distance regression with two parameters, $H_0$ = 73 km/s/Mpc
and  $M_{LG} = 1.26\cdot10^{12} M_{\sun}$, derived as the least-square
solution. The two thin solid lines correspond to the 95\% confidence interval
of the regression. The dashed straight line is an asymptotic Hubble relation
with $H_0$ = 73 km/s/Mpc. The seven probable members of the CVn cloud are
shown as asterisks, and two galaxies, ScuDIG and NGC 247, with poorly
determined distances are marked as triangles.}
%\end{figure*}

%\clearpage
%\begin{figure*}
%\centering
%\includegraphics[width=16.0cm]{Fig7_1.ps}
%\vspace{5mm}
\caption{ The total view of the Local Group and its vicinity in the Cartesian
Supergalactic coordinates. The upper panel shows the Local complex of
galaxies seen in the Supergalactic plane. The nearby galaxies are represented
as "iluminated balls", whose size is proportional to the galaxy luminosity.
Spiral and irregular galaxies are indicated as dark balls, and elliptical
and spheroidal ones are shown as light balls.
The bottom panel presents the Local complex from another point of view.
The Canes Venatici cloud is situated on the right side, and the Sculptor
group is located on the left side. The Local Minivoid occupies the bottom
left corner, and the Local Void of Tully extends above the complex plane
in the "+Z"-direction.}
%\end{figure*}
%\begin{figure*}
%\centering
%\vspace{10mm}
%\hspace{10mm}
%\includegraphics[width=14.0cm]{Fig7_2.ps}
%\setcounter{figure}{4}
%\vspace{5mm}
%\caption{ continued}
%\end{figure*}

%\begin{figure*}
%\centering
%\includegraphics[width=11.0cm,angle=-90]{Fig8.ps}
%\vspace{5mm}
\caption{ The luminosity function of 38 nearby field galaxies from Table 2
(upper histogram) and 96 galaxies situated in the LG, the M81 group,
and the CenA group. E and dSph galaxies are shaded.}
\end{figure}

\end{document}